# Transfer Learning U-Net Deep Learning for Lung Ultrasound Segmentation

Dorothy Cheng[1], Edmund Y. Lam[2]

[1]Department of Bioengineering, Imperial College London. Correspondence.
`tin.cheng19@imperial.ac.uk`
[2]Department of Electrical and Electronic Engineering, The University of Hong Kong

## Abstract

Transfer learning (TL) for medical image segmentation helps deep learning models achieve more accurate performances when there are scarce medical images. This study focuses on completing segmentation of the ribs from lung ultrasound images (LUS) and finding the best TL technique with U-Net, a convolutional neural network (CNN) for precise and fast image segmentation. Two approaches of TL were used, using a pre-trained VGG16 model to build the U-Net (V-Unet) and pre-training U-Net network with grayscale natural salient object dataset (X-Unet). Visual results and dice coefficients (DICE) of the models were compared. X-Unet showed more accurate and artifact-free visual performances on the actual mask prediction, despite its lower DICE than V-Unet. A partial-frozen network fine-tuning (FT) technique was also applied to X-Unet to compare results between different FT strategies, which FT all layers slightly outperformed freezing part of the network. The effect of dataset sizes was also evaluated, showing the importance of the combination between TL and data augmentation (DA).

## 1. Introduction

The emerging development in computer vision technology using deep CNN for medical image segmentation has been considered a great contribution to medical care recently. Image segmentation is a process of dividing an input image into multiple sets of pixels with the same nature to extract the targeted area people are interested in, transforming the medical image into a meaningful subject for diagnostic processes. [1-3] In this paper, we aimed to complete segmentation for LUS using U-Net network structure and find out the best transfer learning technique.

Lung ultrasound has become an important non-invasive assessment tool especially from the 2019 novel coronavirus disease (COVID-19) for lung condition detection and diagnosis such as consolidations, pneumothorax, pleural effusion, etc. [4] The rib in LUS is our targeted region in this study, which is the pivotal landmark exhibiting pronounced acoustic features. The rib appears hyperechoic with acoustic shadows posterior to them. [23] This study also covers the basic implementation of homomorphic wavelet transform to remove speckle noise (SN) in raw LUS which causes resolution degradation.

We chose U-Net as the network architecture framework. It has a CNN structure that is first designed and outperformed in medical image segmentation tasks. [6] It is an end-to-end, pixel-to-pixel fully convolutional network, allowing efficient whole-image-at-a-time learning and dense prediction for per-pixel semantic segmentation. [7] Figure 1 shows its encoder-decoder structure with a series of skip connections. The encoder down-samples and gradually reduces the spatial dimension through merging the layers for feature extraction, and the decoder restores the image details and spatial dimension through up-sampling, giving U-Net the power to seamlessly segment arbitrarily large images with an overlap-tile strategy. [8]

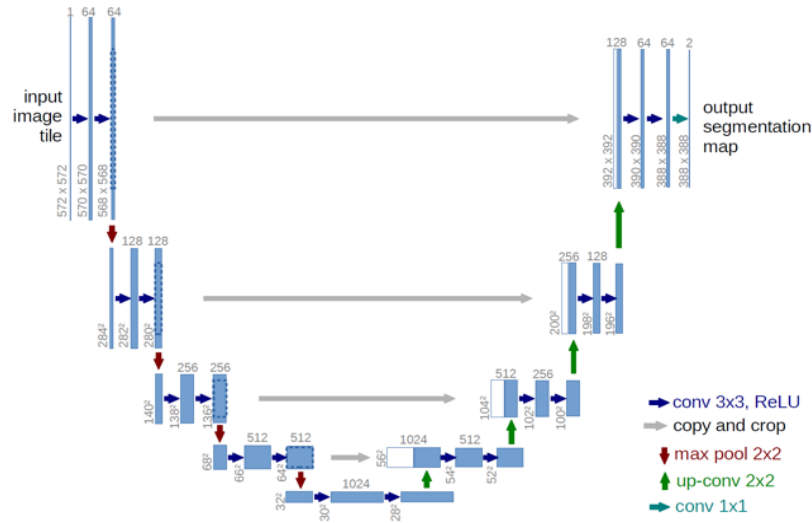

Figure 1: Original U-Net structure for biomedical image segmentation

Transfer learning is another noteworthy technique implemented in this study. It is based on implementing previously learnt knowledge to solve new similar problems in a more effective and efficient manner. TL allows the model to be trained on a smaller amount of data, making it particularly useful in medical image segmentation. [9] One common application of TL in classification and segmentation problems is the use of CNN models pre-trained on ImageNet, such as VGGNet, ResNet, AlexNet and Inception etc. [10] We took this idea and implemented it on U-Net. ImageNet contains 100,000 RGB natural images for solving a range of tasks. Recent research suggested the idea of pretraining U-Net with a grayscale salient object dataset which is more specific for segmentations in general. [13-14] We combined the idea of TL into U-Net with two approaches mentioned above in our LUS segmentation and evaluated their performances.

## 2. Related Work

### 2.1. U-Net for Medical Image Segmentation

U-Net has been widely used in medical image segmentation nowadays, which was first developed in 2015 for microscopy image segmentation of cells, aiming for fast and precise segmentation using a scarce amount of training images. [6] In 2020, U-Net-related papers covered a variety of medical image modalities, with magnetic resonance imaging (MRI) and computed tomography (CT) being the most common forms, and the brain and pathology being the most common application areas. [15-16] U-Net also varies in different forms, such as applying 3D convolutional (conv) layers and inception modules to accommodate specific image modalities and applications. [3, 12, 15, 17-18] Our 2D LUS require only 2D U-Net, with our research focusing mainly on TL techniques. Therefore, the basic 2D U-Net was used. There were a few research implementing basic U-Net particularly on US segmentation, such as article [19] for segmenting the fetal lung and heart and the Ultrasound Nerve Segmentation Kaggle Challenge in 2016. [20] It provided the dataset and U-Net coding examples of segmenting the Brachial Plexus, a collection of nerves, in US images.

### 2.2. Transfer Learning Techniques

The above research only focuses on the structural variants of U-Net. To optimize the model performance on a limited number of medical images, TL helps address the issue. TL is categorized into cross-domain and cross-modal, depending on the domains of the target and source data. Cross-domain TL was used in this study, which is a popular approach to pre-train the model in large non-US datasets such as ImageNet. It is efficient to learn about detections of edges, shapes, and textures from natural images, then modify and fine-tune the model with US images. [10]

Using pre-trained models for TL is a common practice. According to the overview research on TL in breast ultrasound imaging (BUS) for breast cancer diagnosis, the most two common pre-trained models used on BUS were VGGNet followed by AlexNet, especially in classification problems. [10, 21] VGGNet (VGG16 and VGG19) outperformed the latter by superseding large kernel-sized filters with various small kernel-sized filers. VGG16 was chosen as our U-Net encoder due to its fewer conv layers between max-pooling layers, which matches better with the original U-Net structure. This idea was applied to US and MRI images before [11, 22], in which article [22] is a similar work to our study. It used VGG16 as the encoder of U-Net for intravascular ultrasound (IVUS) segmentation of the lumen and media. It compared three training model approaches, simple U-Net, VGG16-UNet without DA, and VGG16-UNet with DA. Its DICE and visual performances are shown in Figure 8, showing the significant improvements from using VGG16 for TL.

Besides the above TL method, recent research on U-Net US segmentation demonstrated the idea of implementing TL by pre-training the U-Net model ourselves. [13] It pre-trained the U-Net with a large image dataset of grayscale natural salient objects to better mimic US images, further on FT the model with simulated US (SUS), BUS, Fetal head US (FUS), and chest X-ray with different layer-freezing approaches and dataset sizes. Results showed that the best two FT strategies were training the whole network and whole network except the bottleneck block (BB) at the bottom. The best DICEs varied from 0.785 in BUS and 0.834 in SUS, also shown in Figure 8, to 0.972 in FUS and 0.98 in chest X-ray. A similar approach was implemented to our LUS segmentation, with tests on different dataset sizes and FT strategies.

## 3. Methodology

### 3.1. Network Architecture

We adopted the basic U-Net architecture framework similar to the one from the original article [6]. TL with FT strategies was applied in two training models, the V-Unet and the X-Unet.

#### 3.1.1. (V-Unet) U-Net with VGG16 as the encoder and further trained on LUS

In V-Unet, as shown in Figure 2, VGG16 was chosen to replace the contracting path of U-Net as a hybrid between these structures, due to its similarity to U-Net's contracting path with a smaller number of parameters. [23] Also, the VGG16 already has weights pre-trained by ImageNet that can be easily accessed and applied to our model, which shortens the training time. [6, 24]

To resemble the VGG16 into a symmetrical U-Net structure, the last three fully connected layers were excluded and new layers for the expanding path were added. The expanding path worked reversely to the VGG16, with up-sampling through deconvolution by transposed conv layers. Concatenated skip connections were used to connect blocks of the same filter size from the contracting path to expanding path to reuse the features and retain more information from previous layers. A $1\times1$ conv layer with sigmoid activation was used as the last layer to map the feature vector from 0 to 1. Due to the binary nature of the feature masks, a threshold value of 0.5 was used to convert all pixels with values above 0.5 to 1, and pixels with values below 0.5 to 0. The whole network had 28,804,545 parameters in total.

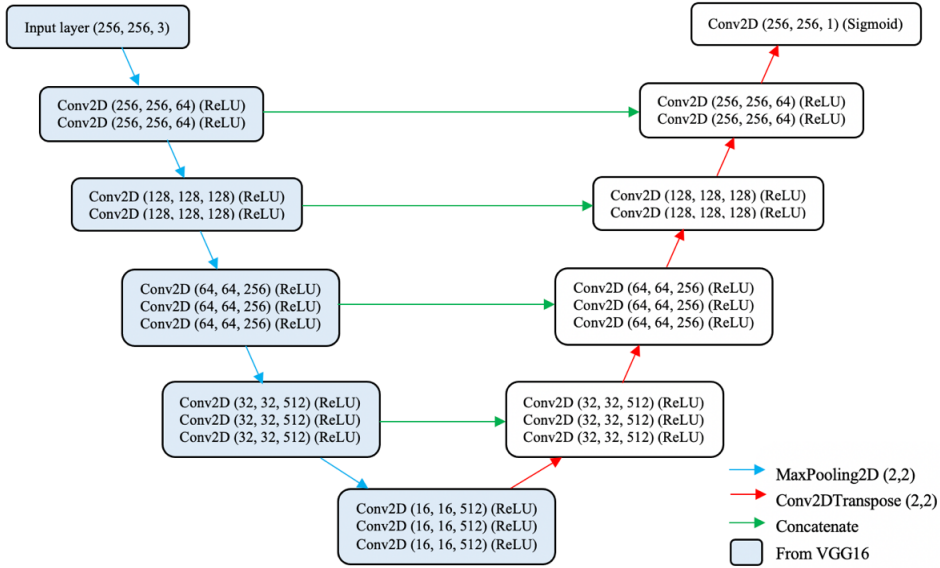

Figure 2: The architecture of V-Unet model

### 3.1.2. (X-Unet) U-Net pre-trained on salient object dataset and fine-tuned on LUS

In X-Unet, as shown in Figure 3, we took the idea of pre-training our own U-Net with grayscale salient object images instead of the publicly available pre-trained models. The network consisted of two 3×3 conv layers followed by ReLU activation in each block. In the contracting path, the number of filters in each block was increased by a factor of 2, starting from 64 filters in the first block to 1024 filters in the fifth block, with 2×2 pixel-window max-pooling layers between blocks. The expanding path, concatenation, and the last 1×1 conv layer were the same as V-Unet's. The whole network had 31,031,745 parameters in total.

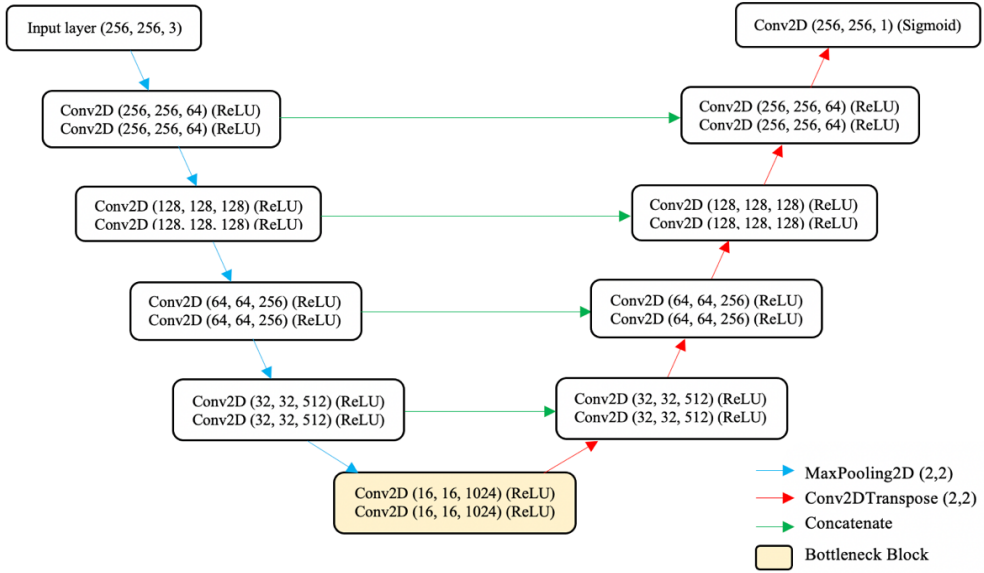

Figure 3: The architecture of X-Unet model

### 3.2. Experimental Design

The difference in network architectures of the two approaches required slightly different TL processes. For V-Unet, the contracting path (VGG16) was frozen in TL, to avoid modifying the ImageNet-pre-trained weights and destroying them in pre-training, as well as to decrease the computation time. [25] LUS were fed into the model for TL. For X-Unet, since no weights were pre-trained before, all layers were not frozen in TL. XPIE dataset was used to pre-train the model. [26]

Upon TL, two models were fine-tuned separately using LUS. We first fine-tuned the whole network, with all layers being trainable (unfrozen), in both V-Unet and X-Unet. An additional FT scheme, training the whole network except the bottleneck block, was implemented on X-Unet and compared its performance to other training models. The BB contains 14,157,824 parameters, which is about half of the parameters of the whole network. Freezing this block has already frozen nearly half of the network, aiming to reduce the training time and maintain accuracy at the same time.

For all the FT processes, five-fold cross-validation was used to assess model performances and provide a range of evaluation scores across different shuffles. 80% of the dataset was used for training, with the remaining 20% for validation.

To also investigate the effects of dataset sizes on the models, experiments on the two U-Net structures with different dataset sizes (200 and 600) were performed. It was aimed to find out the optimal dataset size that can enhance model performance under a relatively small dataset condition since medical images are in general very scarce in availability.

### 3.3. Data Set

#### 3.3.1. LUS Pre-processing

Our LUS dataset consists of raw B-mode images collected from a healthy volunteer (a 26-year-old male). A Verasonics Vantage 256 system was used with a linear array probe operating at 5.2 MHz. Raw LUS were obtained with 960×128 in size stored in MAT format. 200 frames were randomly selected for building our LUS dataset and were annotated manually on MATLAB. Raw US images exhibit speckle noise due to the interference between coherent constructive and destructive energies of scattered echoes, resulting in resolution degradation and making the information extraction process complicated. At the pre-processing stage, homomorphic wavelet transform was applied to LUS for SN reduction. As SN in US images is a type of multiplicative noise, applying logarithmic transform using the following equality can convert SN into addictive noise and be removed by wavelet transform. [27-29]

$$log\, f(x,y) = log\, g(x,y) + log\, \eta_m(x,y) \quad (1)$$

In which $f(x,y)$ represents the real noisy image, $g(x,y)$ represents the unknown noise-free image and $\eta_m(x,y)$ represents the multiplicative noise function. Denoised images were converted into TIFF format.

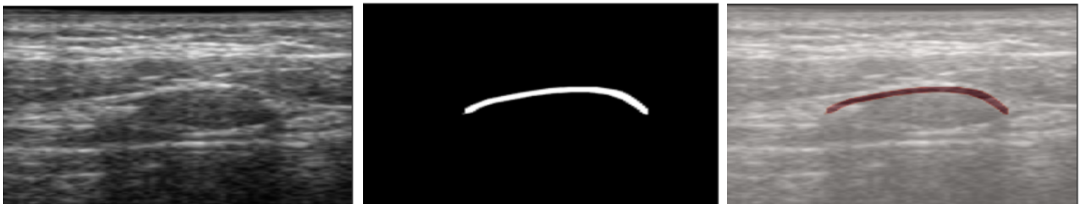

Figure 4: Example of an LUS, binary mask, and the overlapping view

#### 3.3.2. Data Augmentation

DA techniques were implemented to create more data samples from our small LUS dataset and improve network performance. The dataset size was first doubled to 400 images by saving a copy of their horizontally flipped version. Real-time batch-wise DA was then implemented during the training process to ensure the model receives new variations at each epoch. [30] Variations include small-degree random deformation like rotation, width and height shift, shearing, zooming and horizontal flipping. [31] All LUS were resized to 256×256 pixels and normalized to between 0 and 1 to allow faster convergence during training.

### 3.3.3. XPIE dataset

XPIE dataset was used to pre-train the X-Unet for TL purposes. It contains 10,000 segmented natural images originally for salient object detection. [26] The images were originally in RGB structure and 375×500 pixels in size. Article [32] suggested that pre-training on grayscale natural images better mimics the grayscale structure of our single-channeled LUS, leading to higher training accuracy and speed of inference. XPIE images were all converted into grayscale and resized into 256×256 pixels. Same DA strategies were performed in pre-training.

### 3.4. Performance Metrics

The Dice coefficient was chosen as the evaluation metrics. It is a spatial overlap index ranging from 0 to 1, indicating from no spatial overlap to complete overlap between the binary predicted mask and the ground truth. [33] It is a statistical tool that can represent the similarity between two sets of data clearly. [34] The equation is as follows:

$$DICE(f, x, y) = \frac{2 \times \Sigma_{ij} f(x)_{ij} \times y_{ij} + \epsilon}{\Sigma_{ij} f(x)_{ij} + \Sigma_{ij} y_{ij} + \epsilon} \quad (2)$$

In which $x$ is the input image, $y$ is the actual ground truth and $f(x)$ is the prediction output from the model. $\epsilon$ is a small number added to avoid division by zero. [35]

## 4. Results

All experimental models were trained with GPU implementations. NVIDIA Tesla T4 provided by Google Colaboratory and NVIDIA GeForce GTX 1080 Ti were used to accelerate the forward propagation and backpropagation routines. [36-37] All the proposed models were implemented with TensorFlow v2.6.0, with the use of Keras library. [38]

Both TL and FT were performed using the Adaptive Moment Estimation (ADAM) optimizer. Different learning rates (LR) were tested initially and an LR of $10^{-5}$ in TL and $5^{-6}$ in FT were used in our experiments that gave the following results. Using a smaller LR for FT allows smaller changes made to the weights in each update for detailing the training.

Table 1 shows all the DICEs from both V-Unet and X-Unet with different training conditions and strategies. For TL in both models, five-fold cross-validation was not implemented so only one (highest) DICE was shown, while the highest and average DICE were shown for the FT models.

### 4.1. Comparing V-Unet and X-Unet trained with 400 LUS

By comparing the DICEs, V-Unet has the highest of 0.8632, which is 0.034 and 0.053 higher than X-Unet FT with all layers (0.8297) and FT with BB frozen (0.8107) respectively.

However, the visual results are surprisingly different from their DICE performances. Figure 5 shows the comparison of five visual examples between the original mask and four of our models with different training conditions. All the models were able to predict the general shape and location of the rib from LUS accurately. However, both V-Unet TL and FT models were obviously susceptible to artifacts and outliers, despite there were significant improvements in outlier elimination and refinement after FT. Predictions from V-Unet particularly had rough edges and sharp corners (figures 5d-e) that didn't exist in the ground truth. On the contrary, even though both X-Unet models had significantly lower DICEs, their predictions had an overall better performance in shaping and edge smoothing, as well as greater resistance to outliers than V-Unet.

To compare the two different FT techniques in X-Unet, although FT with all layers performed slightly better in both DICE and visual results than FT with BB frozen, the visual differences were insignificant. However, FT with BB frozen did not eventually cause a huge reduction in training time. FT with all layers would be a preferred strategy in this investigation.

| Model | V-Unet | | | | X-Unet | | | |
|---|---|---|---|---|---|---|---|---|
| Training Condition | With 400 LUS | | With 200 LUS | | TL (With XPIE) | FT With 400 LUS | | FT With 600 LUS |
| | TL | FT | TL | FT | | All layers (ALL) | All layers except bottleneck block (BBFrozen) | All layers |
| Highest | 0.8200 | ***0.8632*** (Fold 4) | 0.7424 | 0.7815 (Fold 4) | 0.7189 | ***0.8297*** (Fold 2) | 0.8107 (Fold 2) | 0.8361 (Fold 5) |
| Average | N/A | 0.8603 | N/A | 0.7787 | N/A | 0.8179 | 0.8097 | 0.8329 |

Table 1: DICEs of all the V-Unet and X-Unet models with different training conditions and dataset sizes. X-Unet TL was trained with natural images without LUS, which cannot be used directly to predict actual LUS. DICE is listed as a reference without direct comparing value.

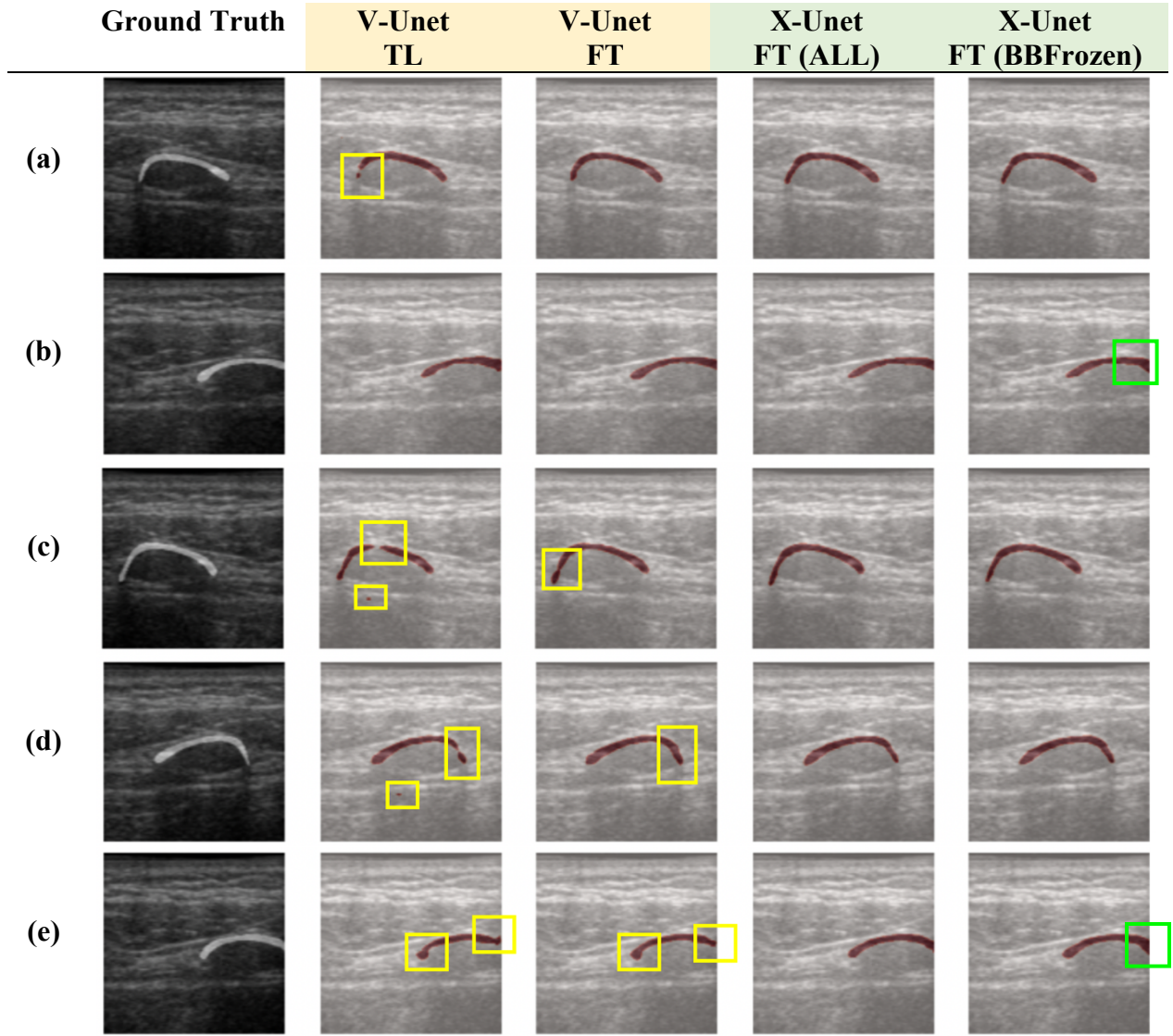

Figure 5: Comparison of five examples (a-e) from the LUS data between the original mask, predicted masks from V-Unet TL, V-Unet FT, X-Unet FT (All layers), and X-Unet FT (BBFrozen). All of them were trained with 400 LUS. Yellow boxes indicate the defects from V-Unet. Green boxes indicate the defects from X-Unet.

## 4.2. Comparing V-Unet and X-Unet trained with different dataset sizes

Firstly, in the V-Unet, we trained with the same strategy with 200 LUS (without the horizontally flipped copies) and compared its FT model with the performance of that trained with 400 LUS. Results in Table 1 show that training with 200 LUS had a DICE 0.082 lower than that trained with doubled LUS, which is also consistent with the visual results in Figure 6. The one trained with 400 LUS showed much fewer artifacts than that with 200 LUS especially on the two ends of the rib. Training with 200 LUS made the results prone to outliers.

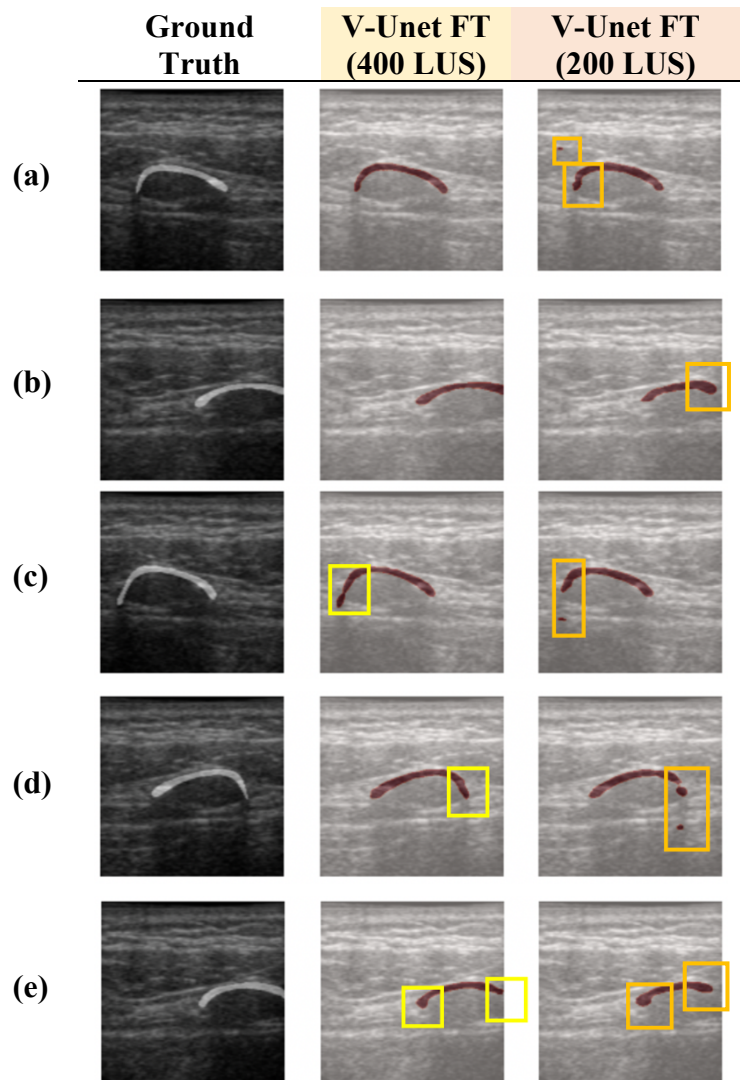

Figure 6: Comparison of 5 examples (a-e) from the LUS data between the original mask, predicted masks from V-Unet TL model trained with 400 LUS and 200 LUS. Yellow boxes indicate the defects of V-Unet (400). Orange boxes indicate the defects of V-Unet (200).

To justify whether a larger dataset size can significantly improve the model accuracy, extra 100 random frames of LUS with their horizontally flipped copies were added to the dataset to create a total of 600 images for FT all layers in X-Unet. The DICE and visualized results are also shown in Table 1 and Figure 7. It shows that there is less than 0.01 increase in DICE when the dataset is 1.5 times larger. Both models produced similar masks except for (7c) and (7e) which there was a slight improvement in the length at the right end. All in all, increasing the dataset size indeed gives a higher DICE but the visual improvements are insignificant in this case.

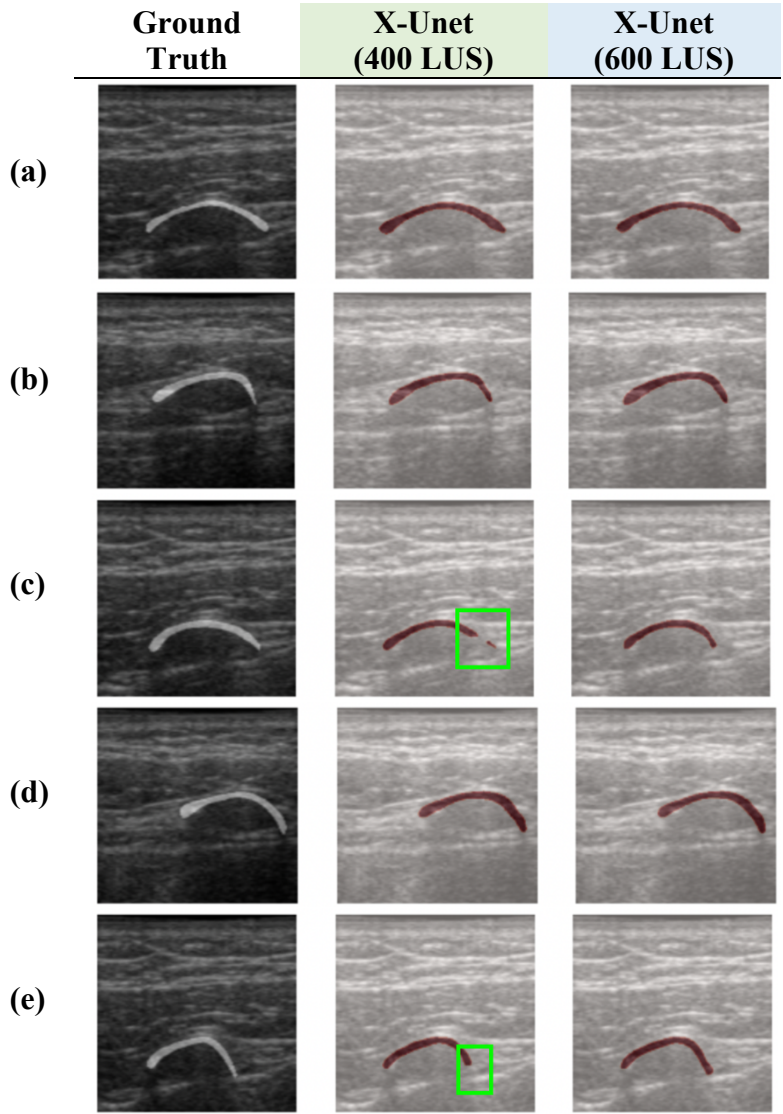

Figure 7: Comparison of 5 examples (a-e) from the LUS data between the original mask, predicted masks from X-Unet FT model trained with 400 LUS and 600 LUS. Green boxes indicate the defects of X-Unet (400).

## 5. Discussion

Both V-Unet and X-Unet exhibited the importance of TL and FT by having significant improvements in DICEs after FT which can also be visualized from TL and FT model prediction comparison of V-Unet. However, V-Unet models are generally very vulnerable to artifacts and outliers that cannot be reflected from its high DICE. It may be due to the fact that an LR of $10^{-5}$ for TL is too large in V-Unet, causing the model to converge too quickly without fully learning and adapting to the edge details of LUS. Future investigation can implement adaptive LR through callbacks during training, which the model can monitor and reduce LR when DICE has stopped improving for a certain amount of epoch, aiming to fine-tune better on model weights and obtain a fit learning schedule. [39-40]

The approach suggested in [13] to pre-train the model with the grayscale XPIE dataset was also demonstrated. Predictions from X-Unet were the closest to the ground truth among all models, despite its lower DICE to V-Unet. The XPIE dataset targets salient objects, which have more attentive edges and regions. Despite the resource- and time-consuming pre-training procedure, X-Unet models perform surprisingly better in edge smoothing, region shaping, and resisting artifacts and outliers especially at the two ends of the segmented rib. Regarding the additional FT technique of freezing the BB in X-Unet, although the DICE wasn't as high as expected without a reduction in training time, the visual outcome was still comparable and considered successful. The reason behind the reduced DICE is because there is still a huge dissimilarity between the XPIE dataset and LUS. Freezing the deeper layers like the BB would make the network not specific enough in extracting special features in LUS, hence lowering the training accuracy.

DICE, our evaluation metric, was achieved on a satisfactory level numerically greater than 0.8. However, it cannot achieve greater than 0.9 due to the fluctuation in the ability to resist outliers and artifacts, as shown from our model predictions. It may be due to the subjectivity and inconsistency from the manual annotation of the ground truth from a non-medical professional. There is subjective boundary uncertainty due to individual skills and image qualities. [41] Further investigation on this should consult medical professionals for annotation.

To evaluate whether DICE is a suitable evaluation metric, there was consistency on visual results within V-Unet or X-Unet model itself. The increase in DICE from TL to FT within the same model matches with the improved visual accuracy, which may not always be the case. Figure 8 shows the visual results and DICEs from articles [22] and [13] that have model structures similar to our V-Unet and X-Unet respectively. For media segmentation indicated by the red box, DICE achieved 0.9771 while the mask was obviously smaller than the ground truth, which was a worse prediction than that with DICE of 0.9257. We were able to produce consistently improving predictions with increasing DICEs within the model itself.

However, DICE is not a suitable indicator to directly compare performances between different models. As shown in section 4, DICEs of V-Unet FT model were considerably higher than that of X-Unet, while visual results from X-Unet significantly outperformed the former in terms of accuracy, shapes and edges. This inconsistency between models is also demonstrated in Figure 8, where both lumen segmentation from [22] and SUS segmentation from [13] produced visual predictions very close to their ground truths, while they have a difference of 0.13 in DICE. DICE cannot be used alone as an evaluation indicator for different model comparisons without the interpretation of visual results. Other metrics can also be explored in the future as an extra reference to model evaluation.

| Similar to V-Unet: | **Original Image** | **Ground Truth** | **Simple U-Net** | **VGG16-UNet with DA** |
|---|---|---|---|---|
| **Lumen Seg in [22]** | 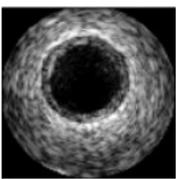 | 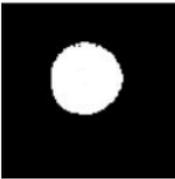 | 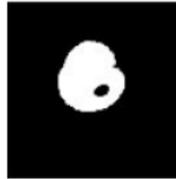 DICE: 0.8041 | 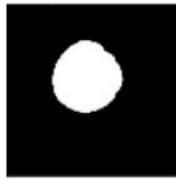 DICE: 0.9629 |
| **Media Seg in [22]** | 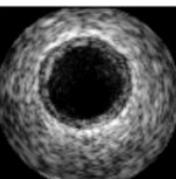 | 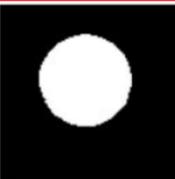 | 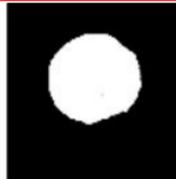 DICE: 0.9257 | 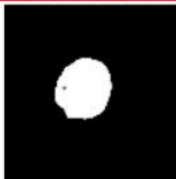 DICE: 0.9771 |
| Similar to X-Unet: | **Original Image** | **Ground Truth** | **U-Net FT with whole network** | **U-Net FT except BB** |
| **SUS Seg in [13]** | 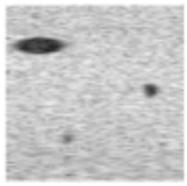 | 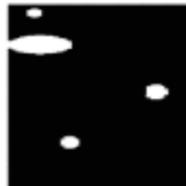 | 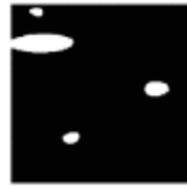<br>DICE: 0.834 | 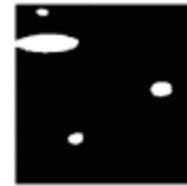<br>DICE: 0.826 |
| **BUS Seg in [13]** | 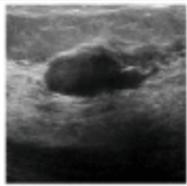 | 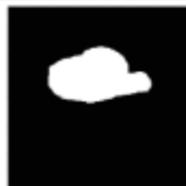 | 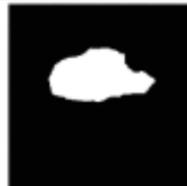<br>DICE: 0.774 | 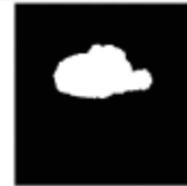<br>DICE: 0.785 |

Figure 8: Visual results and DICEs of IVUS segmentation from [22], SUS and BUS segmentation from [13]

In terms of the effects of dataset sizes, there is a huge deterioration in results with 200 LUS on V-Unet model. This shows the strong need of DA for such a scarce number of images. Additionally, we expanded the dataset to 1.5 times larger particularly for testing whether it would give a significant improvement. X-Unet model just showed a slight increase of accuracy in both visual results and DICE. It is consistent with the fact that a larger dataset leads to better model performance from creating larger variance, but the level of improvement is not proportional. To aim at better results, further investigation on network structures or parameters is emphasized to compromise with the limited amount of data available.

There are also other suggested improvements that could be done in the future, such as exploring other freezing techniques in the FT process. In this study, the main FT strategy used is training the whole network without freezing any layer. It is one of the easiest strategies to yield optimal results, but it is time-consuming for all the parameters to be updated. More of the FT strategies such as those suggested in [13] could be tried out in the future.

It is also noteworthy to investigate on network regularization. Our network structures were based on the architecture suggested in [6] without regularization layers. One of the potential drawbacks of our models is that they may be over-fitting to US images only. Regarding this, batch normalization (BN) layers were implemented after conv layers into our V-Unet during the experimental period as a trial. However, training did not converge and the model failed to make predictions. It was because our batch size used (20) was small and our LUS were all similar without large fluctuation. They weren't in a need of normalization within batches and adding BN might, in turn, give a noisy approximation. [42] Further investigation on implementing other regularizations in the network for improving the model performance and preventing over-fitting is suggested.

## 6. Conclusion

This study presented the V-Unet and the X-Unet with the implementation of TL into U-Net for completing the LUS segmentation of ribs. V-Unet used VGG16 as the encoder of the U-Net, while X-Unet was pre-trained with grayscale natural salient object images. Although V-Unet achieved higher DICE than X-Unet, X-Unet outperformed the V-Unet in visual predictions with a significant refinement in shaping and edge smoothing. The evaluation must be done with both DICE and visual interpretations since direct comparison of DICE between different models is unreliable. We also performed a brief investigation on using different FT strategies and dataset sizes on model predictions. It is suggested that future investigation on monitoring the LR, modifying the network structure details and FT techniques are essential to achieve more accurate predictions.

## 7. Acknowledgement

We acknowledge Dr. Wei-Ning Lee and Xiao Fei Sun from the Department of Electrical and Electronic Engineering, The University of Hong Kong, for providing the lung ultrasound images and guidance on ultrasound image interpretations.

## 9. Abbreviations

| | |
|---|---|
| ADAM | Adaptive Moment Estimation |
| BB | Bottleneck Block |
| BN | Batch Normalization |
| BUS | Breast Ultrasound |
| conv | Convolutional |
| CNN | Convolutional Neural Networks |
| CT | Computed Tomography |
| DA | Data Augmentation |
| DICE | Dice Coefficient |
| FUS | Fetal Head Ultrasound |
| FT | Fine-tuning |
| IOU | Intersection Over Union |
| IVUS | Intravascular Ultrasound |
| LUS | Lung Ultrasound Imaging |
| LR | Learning Rate |
| MRI | Magnetic Resonance Imaging |
| RGB | Red Green Blue |
| SN | Speckle Noise |
| SUS | Simulated Ultrasound |
| TL | Transfer Learning |
| US | Ultrasound |